\documentstyle[12pt,axodraw]{article}
\begin{document}
\thicklines
\thispagestyle{empty}
\begin{flushright}
BI-TP~~94/61
\end{flushright}
\vspace*{15mm}
%
\font\tenbf=cmbx10
\font\tenrm=cmr10
\font\tenit=cmti10
\font\elevenbf=cmbx10 scaled\magstep 1
\font\elevenrm=cmr10 scaled\magstep 1
\font\elevenit=cmti10 scaled\magstep 1
\newcommand{\Sp}[1]{{\mbox{Li}}_2\left(#1\right)}
\newcommand{\Cl}[1]{{\mbox{Cl}}_2\left(#1\right)}
\newcommand{\rd}{{\rm d}}
\newcommand{\ep}{\varepsilon}
\newcommand{\Pm}{\phantom-}
\newcommand{\Pu}{\phantom1}
\newcommand{\Df}[2]{\mbox{$\frac{#1}{#2}$}}
\newcommand{\Pp}[3]{#1.#2\!\times\!10^{-#3}}
\newcommand{\Fh}[2]{\,{}_#1F_#2}
\newcommand{\Fs}[3]{\!\!\left[\begin{array}{c}#1\,;\\#2\,;\end{array}#3\right]}
\newcommand{\Fu}[2]{\Fs{#1}{#2}{1}}
\newcommand{\Ff}[2]{\Fs{#1}{#2}{4}}
\newcommand{\Fq}[2]{\Fs{#1}{#2}{\frac{1}{4}}}
\newcommand{\Fuq}[2]{\Fs{#1}{#2}{\frac{-q^2}{m^2}}}
\newcommand{\Fum}[2]{\Fs{#1}{#2}{\frac{m^2}{-q^2}}}
\newcommand{\Ffq}[2]{\Fs{#1}{#2}{\frac{-q^2}{4m^2}}}
\newcommand{\Ffm}[2]{\Fs{#1}{#2}{\frac{4m^2}{-q^2}}}
\newcommand{\Ffz}[2]{\Fs{#1}{#2}{z}}
\newcommand{\Fft}[2]{\Fs{#1}{#2}{1-t}}
\newcommand{\Fzz}[2]{\Fs{#1}{#2}{1-z}}
\newcommand{\ba}{\begin{eqnarray}}
\newcommand{\ea}{\end{eqnarray}}
\newcommand{\crn}{\nonumber \\}

\newcommand{\xip}{\xi^\prime(1)}
\newcommand{\Xp}{X^\prime(0)}
\newcommand{\ei}{(\exp({\rm i}\theta))}
\newcommand{\dfr}[2]{\mbox{$\frac#1#2$}}
\parindent=3pc
\baselineskip=10pt
\begin{center}
\def\thefootnote{\fnsymbol{footnote}}
   {{\tenbf ALGORITHMIC CALCULATION OF TWO-LOOP FEYNMAN DIAGRAMS
\footnote
{~to be published in the proceedings of the workshop on Computer Algebra in
Science and Engineering, held at Zentrum f\"ur Interdisziplin\"are
Forschung (ZiF), Bielefeld, Germany, August 1994; World Scientific,
J.Fleischer, J.Grabmeier, F.W. Hehl and W.K\"uchlin, editors.}
 }\\}

\vspace{16mm}

{\large J.~Fleischer,~~~~~~~~~}%
\newcommand{\st}{\fnsymbol{footnote}}%
\medskip
{\large O.~V.~Tarasov \footnote
{~Supported by Bundesministerium f\"ur Forschung und Technologie.}
\footnote
{~On leave of absence from Joint Institute for Nuclear Research,
141980 Dubna, Moscow Region, Russian Federation.}
}

\medskip
{\em   Fakult\"at f\"ur Physik, Universit\"at Bielefeld
		    D-33615 Bielefeld 1, Germany}

\bigskip

\vspace*{20mm}

\textwidth 120mm
\begin{abstract}

\vglue 0.3cm
{\rightskip=3pc
\leftskip=3pc
\tenrm\baselineskip=12pt
\noindent
 In a recent paper \cite{ft}
 a new  powerful method to calculate Feynman diagrams
was proposed. It consists in setting up a
 Taylor series expansion  in the external momenta squared.
The Taylor coefficients are obtained from the original
diagram by differentiation and putting the
external momenta  equal to zero.
It was demonstrated that by a certain conformal mapping and
subsequent resummation by means of Pad\'{e} approximants it is
possible to obtain high precision numerical values
of the Feynman integrals in the whole cut plane.
The real problem in this approach is the calculation
of the Taylor coefficients for the arbitrary mass case.
Since their analytic evaluation by means of CA packages
uses enormous CPU and yields very lengthy expressions,
we develop an algorithm with the aim to set up a
FORTRAN package for their numerical evaluation.
This development is guided by the possibilities offered
by the formulae manipulating language FORM \cite{FORM}.
\vskip 6.0cm
\vglue 0.6cm}

\end{abstract}
\end{center}

\bigskip


\textwidth 170mm
\newpage

\setcounter{page}1

\renewcommand{\thefootnote}{\arabic{footnote}}
\setcounter{footnote}{0}

{\elevenbf\noindent 1. Introduction }
\vglue 0.2cm

 Standard-model radiative corrections of high accuracy
have obtained
growing  attention lately in order to cope with the
increasing precision of LEP experiments \cite{LEP}.  In particular
two-loop calculations with  nonzero masses became
relevant \cite{2loop}. While in the one-loop approach
there exists a systematic way of performing these calculations
\cite{1loop}, in the two-loop case there does not exist such
a developed technology and only  a series of partial
results were obtained \cite{methods}, \cite{asy}
but no systematic approach was formulated.

   Our approach consists essentially in
performing a Taylor series expansion in terms of external momenta squared
and analytic continuation into the whole  region of kinematical
interest. Simple as
this may sound, there are some unexpected methodical advantages compared
to other procedures.

   Considering a Taylor series expansion in terms of one external momentum
squared, $q^2$ say, the differential operator by the repeated application
of which the Taylor coefficients are obtained, subsequently setting
$q = 0$, is
\begin{equation}
    \Box_q = \frac{{\partial}^2}{\partial{q_{\mu}}\partial{q^{\mu}}}.
\end{equation}
Such expansions were considered in \cite{Recur}, Pad\'{e} approximants
were introduced in \cite{bft} and in Ref. \cite{ft} it was demonstrated
that this approach can be used to calculate Feynman diagrams on their
cut which, concerning physics, is the most interesting case. The above
Taylor coefficients are essentially ``bubble diagrams'', i.e. diagrams
with external momenta equal zero.
They are essentially the same
(after partial fraction decomposition) for two-point, three-point,
{\ldots} functions for a given number of loops
and we  stress that it is indeed a great technical
simplification
to have to perform integrals only for external
momenta equal zero  even if these integrals contain now arbitrary high powers
 of the scalar
propagators. For their calculation recurrence relations are quite
effective (\cite {Recur},\cite {Bij/Ve}, see Sect.6).
On this basis we develop our algorithm.

\vglue 0.6cm
{\elevenbf\noindent 2. Expansion of three-point functions in terms of
external momenta squared}
\vglue 0.2cm
Here we have two independent external momenta in $d=4-2 \varepsilon$
dimensions.
The general expansion of (any loop) scalar 3-point function with its
momentum space representation $C(p_1, p_2)$ can be written as
\begin{equation}
\label{eq:exptri}
C(p_1, p_2) = \sum^\infty_{l,m,n=0} a_{lmn} (p^2_1)^l (p^2_2)^m
(p_1 p_2)^n = \sum^\infty_{L=0} \sum_{l+m+n=L} a_{lmn}
(p^2_1)^l (p^2_2)^m (p_1 p_2)^n,
\label{2.2}
\end{equation}
where the coefficients 	$a_{lmn}$ are to be determined from the given diagram.
They are obtained by applying the differential operators
$\Box_{ij} = \frac{\partial}{\partial p_{i\mu}}
\frac{\partial}{\partial p_j^\mu}$
several times to both sides of (\ref{eq:exptri}).

This procedure results in a system of linear equations for the $a_{lmn}$. For
fixed $L$ (see equation (\ref{eq:exptri}))
 we obtain a system of $(L+1)(L+2)/2$ equations
of which, however, maximally $\left[ L/2 \right] + 1$ couple ($\left[
x\right]$ standing
here for the largest integer $\le x$). These linear equations are easily
solved with REDUCE \cite{REDUCE}, e.g., for arbitrary $d$.

For the purpose of demonstrating the method, we confine ourselves to the case
$p^2_1 = p^2_2 = 0$, which is e.g. physically realized in the case of the
Higgs decay into two photons ($H \to \gamma \gamma$) with $p_1$ and $p_2$
the momenta of the photons. In this case only the coefficients $a_{00n}$ are
needed. They are each obtained from a ``maximally coupled" system of
$\left[ n/2\right]
+1$ linear equations. Solving these systems of equations we obtain a sequence
of differential operators ($Df$'s)
which project out from the r.h.s. of (\ref{eq:exptri})
the coefficients $a_{00n}$:
\begin{equation}
Df_{00n}=\sum^{[n/2]+1}_{i=1}\frac{(-4)^{1-i}\Gamma(d/2+n-i)\Gamma(d-1)}
 {2 \Gamma(i) \Gamma(n-2i+3) \Gamma(n+d-2) \Gamma(n+d/2)}(\Box_{12})^{n-2i+2}
 (\Box_{11}\Box_{22})^{i-1},
\label{3.3}
\end{equation}
where the sum of the exponents of the various $ \Box ' $ s is equal $n$.
Applying the operator $Df_{00n}$ to the (scalar) momentum space integral
$C(p_1,p_2)$ and putting the external momenta equal to zero, yields the
expansion coefficients $a_{00n}$.

In the two-loop case we consider the scalar
integral ($k_3 = k_1 - k_2$, see also Fig.~1)
\begin{eqnarray}
\label{treug2}
\begin{array}{l}
C(m_1, \cdots, m_6; p_1, p_2)\\
\\
= \frac{1}{(i\pi^2)^2} \int
\frac{d^4 k_1 d^4 k_2}{((k_1 + p_1)^2 -m^2_1)((k_1 + p_2)^2 - m^2_2)
((k_2 + p_1)^2 - m^2_3) ((k_3 + p_2)^2 - m^2_4) (k^2_2 - m^2_5)
(k^2_3 - m^2_6)},
\end{array}
\label{2.4}
\end{eqnarray}


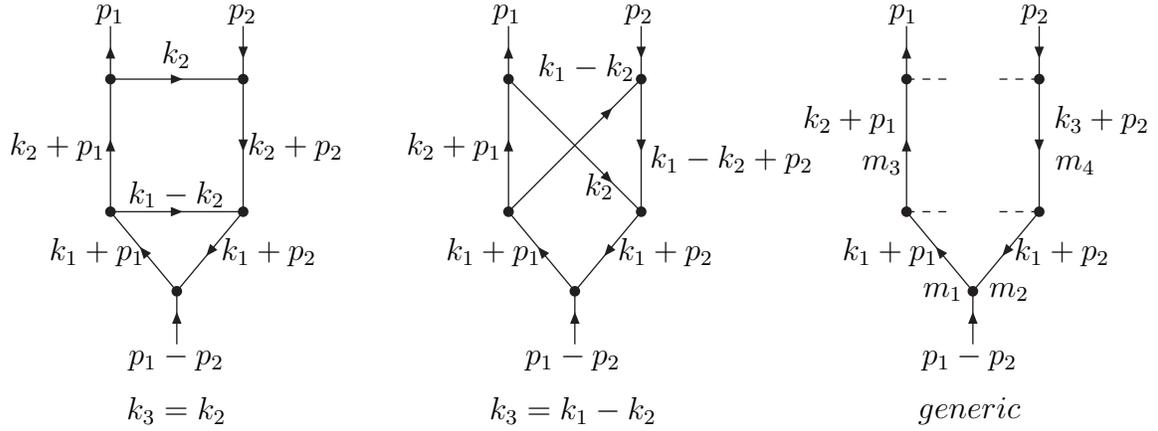
\begin {figure} [htbp]
\begin{picture}(400,190)(-35,0)

\ArrowLine(10,100)(60,100)
\ArrowLine(10,150)(60,150)
\ArrowLine(10,100)(10,150)
\ArrowLine(60,150)(60,100)
\ArrowLine(60,100)(35,70)
\ArrowLine(35,70)(10,100)

\ArrowLine(35,50)(35,70)
\ArrowLine(10,150)(10,170)
\ArrowLine(60,170)(60,150)


\Text(-10,125)[]{$k_2+p_1$}
\Text(80,125)[]{$k_2+p_2$}
\Text(35,160)[]{$k_2$}
\Text(35,45)[]{$p_1-p_2$}
\Text(35,25)[]{$k_3=k_2$}
\Text(5,85)[]{$k_1+p_1$}
\Text(70,85)[]{$k_1+p_2$}
\Text(35,107)[]{$k_1-k_2$}
\Text(10,175)[]{$p_1$}
\Text(60,175)[]{$p_2$}

\Vertex(10,100){2}
\Vertex(60,100){2}
\Vertex(10,150){2}
\Vertex(60,150){2}
\Vertex(35,70){2}

\Line(160,100)(185,125)
\Line(160,150)(185,125)
\ArrowLine(185,125)(210,150)
\ArrowLine(185,125)(210,100)
\ArrowLine(160,100)(160,150)
\ArrowLine(210,150)(210,100)
\ArrowLine(185,70)(160,100)
\ArrowLine(210,100)(185,70)


\ArrowLine(185,50)(185,70)
\ArrowLine(160,150)(160,170)
\ArrowLine(210,170)(210,150)

\Vertex(160,100){2}
\Vertex(210,150){2}
\Vertex(185,70){2}
\Vertex(160,150){2}
\Vertex(210,100){2}

\Text(140,125)[]{$k_2+p_1$}
\Text(245,120)[]{$k_1-k_2+p_2$}
\Text(195,110)[]{$k_2$}
\Text(185,45)[]{$p_1-p_2$}
\Text(185,25)[]{$k_3=k_1-k_2$}
\Text(155,85)[]{$k_1+p_1$}
\Text(220,85)[]{$k_1+p_2$}
\Text(190,155)[]{$k_1-k_2$}
\Text(160,175)[]{$p_1$}
\Text(210,175)[]{$p_2$}


\ArrowLine(310,100)(310,150)
\ArrowLine(360,150)(360,100)
\ArrowLine(335,70)(310,100)
\ArrowLine(360,100)(335,70)


\ArrowLine(310,150)(310,170)
\ArrowLine(360,170)(360,150)
\ArrowLine(335,50)(335,70)
\DashLine(310,150)(325,150){3}
\DashLine(345,150)(360,150){3}
\DashLine(310,100)(325,100){3}
\DashLine(345,100)(360,100){3}

\Vertex(310,100){2}
\Vertex(310,150){2}
\Vertex(360,100){2}
\Vertex(360,150){2}
\Vertex(335,70){2}

\Text(290,135)[]{$k_2+p_1$}
\Text(385,135)[]{$k_3+p_2$}
\Text(335,45)[]{$p_1-p_2$}
\Text(335,25)[]{$generic$}
\Text(305,85)[]{$k_1+p_1$}
\Text(370,85)[]{$k_1+p_2$}
\Text(310,175)[]{$p_1$}
\Text(360,175)[]{$p_2$}

\Text(302,118)[]{$m_3$}
\Text(325, 70)[]{$m_1$}
\Text(375,118)[]{$m_4$}
\Text(350, 70)[]{$m_2$}

\end{picture}
\caption {Planar and non-planar scalar vertex diagrams and their kinematics
}
\end{figure}

Introducing the abbreviations $c_1=k_1^2-m^2_1, c_2=k_1^2-m^2_2,
c_3=k_2^2-m^2_3, c_4=k_3^2-m^2_4$ and $c_5=k_2^2-m^2_5, c_6=k_3^2-m^2_6$,
we have
($c_5$ and $c_6$ do not enter the differentiation since for the planar
as well as for the non-planar diagram they occur in (\ref{treug2}) as such)
\begin{equation}
\label{e24}
(i\pi^2)^2 a_{00n} = \frac{2^n}{n+1} \int d^4 k_1 d^4 k_2 F_n \cdot
\frac{1}{c_1~ c_2~ c_3 ~ c_4~ c_5 ~ c_6},
\end{equation}
where for the {\it planar} diagram ($k_3=k_2$ in $c_4$ and $k_3=k_1-k_2$
in $c_6$)
\begin{equation}
\label{Fn}
F_n = \sum^n_{\nu=0} c_1^{-(n-\nu)} c_3^{-\nu} \sum^n_{\nu^\prime=0}
c_2^{-(n-\nu^\prime)} c_4^{-\nu^\prime} \cdot A^n_{\nu\nu^\prime} (k_1, k_2),
\end{equation}
and
\begin{equation}
\label{hernja}
A^n_{\nu\nu^\prime} (k_1, k_2) = \sum_{0\le 2\mu \le \nu + \nu^\prime \le
n+\mu} a^{n\mu}_{\nu\nu^\prime} (k^2_1)^{n-(\nu + \nu^\prime)+\mu}
(k^2_2)^\mu (k_1~ k_2)^{\nu + \nu^\prime - 2\mu},
\end{equation}
$a^{n\mu}_{\nu\nu^\prime}$ being rational numbers with the properties
\begin{equation}
\label{AAA1}
a^{n\mu}_{\nu\nu^\prime} = a^{n\mu}_{\nu^\prime\nu}	\qquad \mbox{and}
\qquad \sum_\mu a^{n\mu}_{\nu\nu^\prime} = 1,
\end{equation}

   The above is now essentially the basis for the algorithm we are
going to develop: to calculate the Taylor coefficients of integrals
like (\ref{2.4}) according to (\ref{e24}) - (\ref{AAA1}) and to reduce
them to integrals of the type

\begin{equation}
V_B(\alpha,\beta,\gamma,m_1,m_2,m_3)=(-1)^{(\alpha+\beta+\gamma)}\int
\frac{d^dk_1 d^dk_2}{(k_1^2-m_1^2)^{\alpha}(k_2^2-m_2^2)^{\beta}
					   ((k_1-k_2)^2-m_3^2)^{\gamma}} ,
\label{VBs}
\end{equation}
which in turn will be reduced by means of recurrence relations to
$V_B(1,1,1,m_1,m_2,m_3)$. Accordingly our algorithm is performed in
the following three steps:

\begin{itemize}
\item First of all the coefficients $a^{n\mu}_{\nu\nu^\prime}$ in
(\ref{hernja}) and the corresponding ones for the non-planar
diagram can be evaluated in terms of multiple sums over $\Gamma
$-functions. While (\ref{hernja}) has been obtained in \cite{ft} by
inspection of FORM output and some lower coefficients could be read off
explicitly, in Sect.~4 we give a proof of this representation by
construction, which also yields the $a^{n\mu}_{\nu\nu^\prime}$.

\item From (\ref{VBs}) it becomes clear that the numerator scalar
products in (\ref{hernja}) must be eliminated
and/or a partial fraction
decomposition of products of scalar propagators with the same
integration momentum $k_1$, $k_2$ or $k_3$ but different masses
must be performed.
Substituting, e.g., $k_1^2=c_i+m_i^2$ (i=1,2) one cancels $k_1^2$.
Similarly one proceeds for $k_2^2$ and $k_3^2$. For $k_1 k_2$ in
(\ref{hernja}) one writes for the {\it planar} diagram

\begin{equation}
k_1 k_2=\frac12(k_1^2+k_2^2-m_6^2)-\frac12 c_6 \equiv \frac12
k^2-\frac12 c_6
\end{equation}
and by stepwise reducing higher powers
( $ {\nu} + {\nu}^{\prime}-2{\mu} =  {\lambda} $ )
\begin{equation}
\label{kaka}
2^{\lambda} (k_1 k_2)^{\lambda}=(k^2)^{\lambda}-\left[(k^2)^{\lambda-1}+
2 k_1 k_2 (k^2)^{\lambda-2}+ \dots + (2 k_1 k_2)^{\lambda-1}\right] c_6.
\end{equation}

In the second term $c_6$ cancels after insertion into (\ref{e24})
so that only factorized one-loop contributions
are obtained from it (i.e. the integral (\ref{e24}) factorizes into
integrals over $k_1$ and $k_2$ separately).
Moreover, in the square bracket of
(\ref{kaka}) only even powers of $k_1 k_2$ contribute after integration.
The ``genuine" two-loop contributions are then obtained
by replacing $k_1 k_2$ in (\ref{hernja}) by
$ \frac12(k_1^2+k_2^2-m^2_6)$
according to the first term in (\ref{kaka}). The problem of cancelling
$k_1 k_2$ is more complicated for the non-planar diagram, as will be
demonstrated in Sect.~5.

\item The evaluation of the integrals (\ref{VBs}) is supposed to be
performed in terms of recurrence relations, thus reducing them to known
two-loop ``master'' integrals, as will be described in Sect.~6.
If one of the indices $\alpha , \beta , \gamma $ in (\ref{VBs}) is
 $ \leq$ 0 (i.e. in our
notation the corresponding scalar propagator occurs with positive
power in the numerator), the integral can be expressed
again in terms of
factorized one-loop integrals and simple explicit representations can
be found in this case. An example is also given at the end of Sect.~6.

\end{itemize}

\vglue 0.6cm
{\elevenbf\noindent 3. The method of analytic continuation}
\vglue 0.4cm

Before entering the details of the calculation, we want to give
a motivation for the above algorithm, i.e. the main interest in
Feynman diagrams is for the values on their cut and we have to
demonstrate how to obtain these from the Taylor expansion.\\

Assume, the following Taylor expansion of a scalar diagram or a
particular amplitude is given:
\begin{equation}
C(p_1, p_2, \dots) = \sum^\infty_{m=0} a_m y^m \equiv f(y)
\label{orig}
\end{equation}
and the function on the r.h.s.~has a cut for $y \ge y_0$.
In the above case of $H \to \gamma\gamma$ one introduces
$y =\frac{q^2}{4m_t^2}$ with $ q^2 = (p_1 - p_2)^2$ as
adequate variable with $y_0=1$.

 Our proposal for the evaluation
of the original series is in a first step a conformal mapping of the
cut plane into the unit circle and secondly the reexpansion
of the function under consideration
into a power series w.r.to the new conformal variable.
A variable often used \cite{Omega} is
\begin{equation}
\omega=\frac{1-\sqrt{1-\frac{y}{y_0} }}{1+\sqrt{1-\frac{y}{y_0}}}.
\label{omga}
\end{equation}

Considering it as conformal transformation,
the $y$-plane, cut from $y_0$ to $+ \infty$, is mapped into the unit
circle
and the cut itself is mapped on its boundary, the upper
semicircle corresponding to the upper side of the cut.
The origin goes into the point $\omega=0$.\\

After conformal transformation it is suggestive to improve the
convergence
of the new series w.r.to $\omega$ by applying one of the numerous
summation methods \cite{Sha},\cite{Weniger} most suitable for our problem.
We obtained the best results with the Pad\'e method and partially
also with the Levin $v$ transformation. The expansion of
$f(y)$ in terms of $\omega$ is:

\begin{equation}
f(y(\omega))=\sum_{s=0}^{\infty} \omega^{s} \phi_s,
\label{foty}
\end{equation}
where
\begin{eqnarray}
\phi_0&=&a_0 \nonumber \\
&&\nonumber \\
\phi_s&=&\sum_{n=1}^{s}a_n(4y_0)^n\frac{\Gamma(s+n)(-1)^{s-n}}
 {\Gamma(2n) \Gamma(s-n+1)},~~~~~s \geq 1.
\end{eqnarray}

Eq.(\ref{foty}) will be used for the analytic continuation of $f$
into the region of analyticity ($y<y_0$; observe that the series
(\ref{orig}) converges for $\left| y \right|<y_0$ only)
and in particular for the continuation on the cut ($y>y_0$).
In this latter case we write
\begin{equation}
\omega=\exp[i \xi(y)],~~~~~~~~~~~{\rm with}~~~ \cos \xi=-1+2~\frac{y_0}{y}
\end{equation}
and hence

\begin{equation}
f(y)=a_0+\sum_{n=1}^{\infty}\phi_n \exp i n \xi(y)
\label{foncut}
\end{equation}

In any case we have $\left| \omega \right| \leq$ 1 and we will show in the
following how to sum the above series.

Pad\'e approximations are indeed particularly well suited for the summation
of the series under consideration. In the case of two-point functions they
could be shown in several cases (see e.g. \cite{bft} ) to be of
Stieltjes type (i.e. the spectral density is positive). Under this
condition the Pad\'e's of the original series (\ref{orig}) are guaranteed
to converge in the region of analyticity. For the three-point function
under consideration ($H \to \gamma\gamma$), however, the obtained result
shows that the series is not of Stieltjes type (i.e. the
imaginary part changes sign on the cut).

   Having performed the above
 $ \omega $ transformation (\ref{omga}), however, it is rather the
Baker-Gammel-Wills {\it conjecture } ( see \cite{BGW} ), which applies.

 A convenient technique for the evaluation of Pad\'e approximants
is the $\varepsilon$-algorithm
of~\cite{Sha}. In general, given a sequence
$\{S_n|n=0,1,2,\ldots\}$, one constructs a table of approximants using
\begin{equation}
T(m,n)=T(m-2,n+1)+1/\left\{T(m-1,n+1)-T(m-1,n)\right\},
\label{eps}
\end{equation}
with $T(0,n)\equiv S_n$ and $T(-1,n)\equiv 0$. If the sequence $\{S_n\}$ is
obtained by successive truncation of a Taylor series, the approximant
$T(2k,j)$ is identical to the $[k+j/k]$ Pad\'{e} approximant~\cite{Sha},
derived from the first $2k+j+1$ terms in the Taylor series.

 We present results for the
two-loop three-point scalar ( {\it planar}) integral with the
kinematics of the decay $H \rightarrow \gamma \gamma$.
We study the integral (\ref{treug2})  with $m_6=0$ and
all other masses $m_i=m_t (i=1,..,5$). In this special case
all Taylor coefficients can be expressed in terms of
$\Gamma$ -functions. For a list of the first coefficients $a_{00n}$
($\equiv a_n; n=0,...,28)$ see \cite{ft}.

{\scriptsize
\hsize=11in\vsize=8in
\begin{table}[htb]
\caption{Results on the cut $(q^2 > 4m^2_t)$
 in comparison with~\protect\cite{DSZ}. }

\medskip
\def\.{&.}\def\pl{&$\pm$}
\halign{\strut\vrule~\hfil#&#~\vrule
&~\hfil#&#
&~\hfil#&#~\vrule &~\hfil#&# &~\hfil#&#~\vrule	&~\hfil#&# &~\hfil#&#~\vrule
\cr
\noalign{\hrule}
& $q^2/m^2_t$ && [10/10] &&&& [14/14] &&&&  Ref.\protect\cite{DSZ}  && \cr
&&& Re && Im && Re && Im && Re && Im \cr
\noalign{\hrule}  4&.01
&  11&.926 & 12&.66 & 11&.935 & 12&.699 & 11&.9347(1)
& 12&.69675(8) \cr
4&.05
&  5&.195 & 10&.48 & 5&.1952 & 10&.484 & 5&.1952(1) &
10&.4836(4)\cr
4&.10 & 2&.6624 & 9&.095 & 2&.66245 & 9&.0955 & 2&.66246(2)
& 9&.0954(2)\cr
4&.20 & 0&.5161 & 7&.4017 & 0&.516039 &	7&.401640 & 0&.51604(5)
&	7&.40163(4)\cr
4&.50 & - 1&.42315 & 4&.77651 & - 1&.42315097 &
4&.77651003 & - 1&.423122(9) & 4&.776497(9) \cr
5&.0 & - 1&.985805 & 2&.758626 & - 1&.985804823 &
2&.758626375& - 1&.98580(2) & 2&.758625(2)\cr
6&.0 & - 1&.7740540 & 1&.1232494 & - 1&.774053979 &
1&.123249363 & - 1&.77405(1) & 1&.123250(6)\cr
7&.0 & - 1&.4192404 & 0&.4807938 & - 1&.419240377 &
0&.4807938045 & - 1&.419240(5) & 0&.480794(9)\cr
8&.0 & - 1&.13418526 & 0&.1784679 & - 1&.134185262 &
0&.1784687866 & - 1&.134184(1) & 0&.178471(2)\cr
10&.0 & - 0&.75694327 & - 0&.06154833 & -
0&.7569432708 & - 0&.0615483234 & - 0&.756943(1) & - 0&.061547(1)\cr
40&.0 & - 0&.045853 & - 0&.0645673 & - 0&.045852780
& -	0&.0645672604 & - 0&.04585286(7) & - 0&.0645673(9)\cr
400&.0 & + 0&.000082 & - 0&.002167  & + 0&.00008190
&	- 0&.0021670 & 0&.0000818974(3) & - 0&.002167005(3)\cr
\noalign{\hrule}}
\label{pade}
\end{table}
}

Results for this kinematics on the cut are given in Table 1.
The process $H \to \gamma\gamma$ was investigated before in Ref. \cite{DSZ}.
For the master integral under consideration in \cite{DSZ}
all integrations but one could be
performed analytically and only the last one had to be done numerically
(hence the high precision achieved).

Similarly, high precision is obtained on the cut in our approach, as
is demonstrated in Table 1. Here, both real and imaginary
part of the scalar two-loop $H \rightarrow \gamma \gamma$ integral are
shown in  comparison with the results
of Ref. \cite{DSZ}. We consider the
domain $q^2_{thr} <q^2 \leq 100 q^2_{thr}$, where
$q^2_{thr}=4m^2_t ~(m_t=150 GeV)$. For $q^2$ close to the
threshold, the integral has a logarithmic singularity, but
still we obtain good stability of the approximants,
which improves to 8-10 decimals up to $q^2=10q^2_{thr}$
and even for $q^2=100q^2_{thr}$ is still excellent.

 For our methods of analytic
 continuation to work with such high precision,
also the Taylor coefficients must be known with high accuracy.
In general we should know them analytically and then
approximate them with the desired precision.
A good example are the coefficients for the $H \rightarrow \gamma
\gamma$ decay, which can be represented as rational
numbers and which for our purpose were approximated with
a precision of 45 decimals using REDUCE.
In general, however,  it turns out that
the CPU time needed for indices $n {>\atop \sim} 30$ (in (\ref{2.2})
for l = m = 0) is of the order of several hours. Moreover in the
arbitrary mass case the length of the expressions even for lower
indices becomes enormous and is
getting more and more difficult to keep under control.
For these reasons it is not possible to obtain analytic expressions
for the coefficients with a reasonable effort and that is why we
develop a proper algorithm. In the following three sections we
formulate this algorithm according to the items specified in Sect.2.
FORM will be a
guide for the development of the algorithm and a permanent testing
tool by comparing results obtained in different manners.

\newpage

\vskip 1.0cm

\vglue 0.6cm

{\elevenbf\noindent 4. The numerator of the integrand.}
\vglue 0.4cm

The first step to be done is the formal Taylor expansion of the integral
(\ref{2.4}). We choose the following approach: each scalar propagator with an
external momentum $p$ is expanded as ($p^2=0$)
\[
\frac{1}{(k+p)^2 - m^2_i} = \frac{1}{k^2 - m^2_i} \sum^\infty_{j=0}
\left( \frac{-2k p}{k^2 - m^2_i} \right)^j = \frac{1}{c_i} S_i(k,p)\ .
\]
Dealing with the planar and non-planar diagrams (see Fig.~1) simultaneously,
as before, we can express the Taylor coefficients under consideration
in generalization of (5) -- (7) by
\begin{eqnarray}
F_n &=& Df_{00n} S_1(k_1,p_1) S_2(k_1,p_2) S_3(k_2,p_1)
S_4(k_3,p_2)\mid_{p_i=0}\nonumber\\
    &=& \sum^n_{\nu=0} c_1^{-(n-\nu)} c_3^{-\nu} \sum^n_{\nu^\prime=0}
c_2^{-(n-\nu^\prime)} c_4^{-\nu^\prime} \cdot A^n_{\nu\nu^\prime}
(k_1, k_2, k_3)\ .      \label{4.2}
\end{eqnarray}

The differential operator $Df_{00n}$ (\ref{3.3}) applied in (\ref{4.2})
contains two types of operators
\begin{eqnarray*}
\begin{array}{rl}
\bullet & \Box^{n_i}_{12} \ \mbox{with}\ n_i = n-2(i-1) \ \mbox{and}\\
\bullet\bullet & \Box^{i-1}_{11} \Box^{i-1}_{22}
\end{array}
\end{eqnarray*}
such that the sum of the powers $n_i + 2(i-1) = n$. In a first step one has
to find a formula for the application of the $\Box_{12}$ operator. The
result depends on $n_i$, the power to which this operator is raised, and the
partition of scalar products to which it is applied:
\[
tt (n_i, n_i-j_1, n_i - j_2) = \Box^{n_i}_{12} (k_1 p_1)^{j_1} (k_2
p_1)^{n_i - j_1} (k_1 p_2)^{j_2} (k_3 p_2)^{n_i - j_2}\ .
\]

Performing the differentiation with FORM, we find by inspection
\begin{eqnarray*}
tt(l,m,n) &=& \sum^{[\frac{m+n}{2}]}_{j=0} t(l,m,n,j) \cdot
(k^2_1)^{l-(m+n)+j} (k_1 k_2)^{m-j} (k_1 k_3)^{n-j} (k_2 k_3)^j\\
\\
{\rm with} \hspace{1cm} & & \\
\\
t(l,m,n,j) &=& {m \choose       j}      {n \choose j} j! (l - m)! (l - n)!
\frac{l !}{(l-m-n+j)!}\ .
\end{eqnarray*}

Here and in the following we assume inverse powers of factorials of
negative arguments to vanish.
Counting the powers of $p_1$ and $p_2$ in (\ref{4.2}), if we
wish to calculate
$A^n_{\nu\nu^\prime} (k_1, k_2, k_3)$, we need as next
\begin{equation}
2^{(i-1)} (i-1)! dd\left( (n-\nu - j_1) \times k_1, (\nu - n_i + j_1)
 \times k_2
\right)
= \Box^{(i-1)}_{11} (k_1 p_1)^{n-\nu-j_1} (k_2 p_1)^{\nu-n_i+j_1}\ ,
\label{4.5}
\end{equation}
where the function $dd$ of $2(i-1)$ arguments is the totally
symmetric tensor,
contracted with $n-\nu - j_1$ vectors $k_1$ and $\nu - n_i + j_1$ vectors
$k_2$:
\begin{eqnarray*}
dd(m_1, m_2) &=& \delta (m_1, m_2) \\
dd(m_1, m_2, m_3, m_4) &=& \delta (m_1, m_2) \delta(m_3, m_4) + \delta
(m_1, m_3) \delta (m_2, m_4) \\
& & + \delta(m_1, m_4) \delta(m_2, m_3) \qquad \mbox{etc.}\ ,
\end{eqnarray*}
where $\delta$ is Kronecker's delta.
Similarly for the application of $\Box_{22}$ we write
\begin{equation}
2^{(i-1)} (i-1)! dd\left( (n-\nu^\prime - j_2) \times k_1, (\nu^\prime
- n_i + j_2) \times k_2 \right)
= \Box_{22}^{(i-1)} (k_1 p_2)^{n-\nu^\prime - j_2} (k_3 p_2)^{\nu^\prime
-n_i+j_2}\ .    \label{4.6}
\end{equation}

The totally symmetric tensor is implemented in FORM 2.2b and can be used
for performing the differentiation. For high indices $n$, however, in this
manner the differentiation still requires too much time (for $n=32$
appr.~8 hours on the    HP735) and the expressions are too lengthy for
practical use. Therefore the idea is to find an analytic expression for
(\ref{4.5}) and (\ref{4.6}) and thus to obtain finally a formula for the
coefficients $a^{n\mu}_{\nu\nu^\prime}$ in (\ref{hernja}).
Indeed $dd(m \times k_1,(l-m) \times k_2) \equiv dd(m,l-m)$ can be
written as
\begin{equation}
dd(m,l-m)    = \sum^m_{j=P(m), \atop \Delta j =2} d(l,m,j)
(k^2_1)^{\frac{m-j}{2}} (k^2_2)^{\frac{l - m - j}{2}} \ (k_1 k_2)^j
\label{dd}
\end{equation}
with $P(m) = (1-(-1)^m)/2$ and
\[
d(l, m , j)  = \frac{m! (l - m)!}{2^{\frac{l}{2}-j}
\left( \frac{l - m - j}{2} \right) ! \left( \frac{m-j}{2} \right) !
j !}            \ .
\]
(\ref{dd}) has been verified again by inspection of results from
FORM's $dd_-$ ($\ldots$)
function (in FORM 2.2b an algorithm was used but not the evaluation of a
formula).

Summing over all partitions of scalar products, (\ref{4.2}) yields
\begin{eqnarray*}
\lefteqn{A^n_{\nu \nu^\prime} (k_1, k_2, k_3)}\\
& & = (n+1)\ 2^{n-1} \frac{\Gamma(d-1)}{\Gamma(n+d-2)
\Gamma(n+\frac{d}{2})} \sum^{[\frac{n}{2}]+1}_{i=1} (-1)^{i-1}
\frac{(i-1)!}{n_i!} \Gamma (n + \frac{d}{2} - i) f_i
\\
\mbox{with}\\
& & f_i = \sum_{j_1 = \max (0, n_i- \nu)        }^{\min (n_i, n - \nu       )}
	    \sum_{j_2 = \max (0, n_i- \nu^\prime )}^{\min(n_i , n-  \nu^\prime)}
   {n-\nu \choose j_1} {\nu \choose n_i - j_1}
   {n-\nu^\prime \choose j_2} {\nu^\prime \choose n_i - j_2} \\
\\
& &  \cdot tt(n_i, n_i - j_1, n_i - j_2) \cdot dd(n - \nu - j_1, \nu - n_i +
j_1)
\cdot dd(n - \nu^\prime - j_2, \nu^\prime - n_i + j_2) ,
\end{eqnarray*}
where the second $dd$-factor depends on $k_1$ and $k_3$ instead of $k_1$ and
$k_2$ (see (\ref{4.2})).
Rewriting (\ref{dd}) in the following more adequate manner:
\begin{eqnarray*}
\lefteqn{dd(n-\nu-j_1, \nu - n_i + j_1)
= \sum^{[\frac{\nu_j}{2}]}_{\sigma = 0} \Theta(\sigma - [\nu_j - (i-1)])}\\
& & \cdot\      d\Big( 2(i-1),
n-\nu - j_1\ , \nu_j - 2\sigma\Big)
(k^2_1)^{i-1-\nu_j + \sigma}
(k^2_2)^\sigma (k_1 k_2)^{\nu_j - 2\sigma}
\end{eqnarray*}
with $\nu_j = \nu - n_i + j_1$ we obtain
\begin{eqnarray}
\lefteqn{A^n_{\nu\nu^\prime} (k_1, k_2, k_3) = \sum^{\left[
\frac{\nu + \nu^\prime}{2}\right]}_{\mu =0}
\sum^{\left[ \frac{\nu}{2}\right]}_{\sigma =0}
\sum^{\left[ \frac{\nu^\prime}{2}\right]}_{\tau =0}
b^{n\mu,\sigma\tau}_{\nu\nu^\prime}} \nonumber\\
& & \cdot (k^2_1)^{n-(\nu + \nu^\prime)+\mu} (k^2_2)^\sigma (k_1
k_2)^{\nu - \mu - \sigma + \tau}
(k_1 k_3)^{\nu^\prime-\mu+\sigma-\tau}
(k_2 k_3)^{\mu - \sigma - \tau}
(k^2_3)^\tau\ , \label{fullA}
\end{eqnarray}
where the coefficients $b_{\nu\nu^\prime}^{n\mu,\sigma\tau}$ are given by

\newpage

\begin{eqnarray}
b^{n\mu,\sigma\tau}_{\nu\nu^\prime} &=&
(n+1)\ 2^{\lambda - n - 1}\ (n - \nu)!\ (n-\nu^\prime)!
\nu !\ \nu^\prime       !
\frac{\displaystyle\Gamma (d-1)}{\displaystyle\Gamma (n+d-2)
\Gamma (n + \frac{d}{2})} \hfill \\
&& \frac{2^{2\rho}}{\rho ! \sigma ! \tau !}
 \sum_{i=1}^{[ \frac{n}{2} ] + 1} (-4)^{i-1} (i-1)!\
\Gamma (n +
\frac{\displaystyle d}{\displaystyle 2} - i) \hfill \nonumber \\
\nonumber \\
&&\sum_{j=\max (0,n_i-\nu)       }^{\min (n_i, n-\nu)       }   2^j
  \sum_{k=\max (0,n_i-\nu^\prime)}^{\min (n_i, n-\nu^\prime)}   2^k
	 \frac{1}{(n_i - j - \rho)! (n_i - k - \rho)! (\rho + j + k-
n_i)!}\nonumber \\
\nonumber \\
&&
\frac{1}{(\sigma - [j - (n -\nu) + (i-1)])! (\nu_j - 2\sigma)! (\tau - [
k-(n-\nu^\prime) + (i-1)])! (\nu^\prime_k - 2\tau)!} \ , \nonumber
\end{eqnarray}

\medskip

\begin{eqnarray*}
\mbox{with}\qquad \lambda &=& \nu + \nu^\prime - 2\mu\ ,\ \ \rho
= \mu - \sigma - \tau \qquad \mbox{and} \\
\nu_j   &=& \nu - n_i + j\ , \          \nu^\prime_k = \nu^\prime - n_i + k
\ .
\end{eqnarray*}

First of all it is interesting to note that for the
planar diagram (\ref{fullA})
reduces with $k_3 = k_2$ to the simpler form (\ref{hernja}) if we set
\begin{equation}
a^{n\mu}_{\nu\nu^\prime} = \sum^{\left[ \frac{\nu}{2}\right]}_{\sigma =0}
\sum^{\left[ \frac{\nu^\prime}{2}\right]}_{\tau =0}
b^{n\mu,\sigma\tau}_{\nu\nu^\prime} \ , \label{ab}
\end{equation}
i.e. the coefficients $a^{n\mu}_{\nu\nu^\prime}$, which in Ref.~\cite{ft}
(appendix A) were only given explicitly as rational numbers for a limited
number of indices, are now obtained analytically as five-fold sums. The nice
property of the representation (\ref{ab}) is that, once it has been
checked against Ref.~\cite{ft} (which has been done with FORM) for the
planar diagram, the coefficients $b^{n\mu, \sigma\tau}_{\nu\nu^\prime}$
for the nonplanar diagram are checked simultaneously.

\vskip 1.0cm

\vglue 0.6cm

{\elevenbf\noindent 5. Cancellation of the numerator scalar products.}
\vglue 0.4cm

Since we are here interested in developing an algorithm for the calculation
of the Taylor   coefficients, the above is the first necessary step. The next
is to investigate the possibility of cancelling the numerator scalar products
of integration momenta against the bubble propagators $c_i\ (i = 1, \ldots,
4)$. While in a formula manipulating language this is done blindly e.g.~by
using the ``repeat" command of FORM, here we have to find a detailed
prescription if finally our algorithm is to be implemented in terms of
a FORTRAN program. Cancellation of scalar products yields ``genuine" two-loop
bubble integrals which are investigated in terms of recurrence relations in
Sect.~6 and factorized one-loop integrals.

At first we study the {\it planar} diagram.
Due to (\ref{Fn}), (\ref{hernja}) and (\ref{kaka}) we can write
\begin{eqnarray*}
\lefteqn{F_n = \sum_{\nu,\nu^\prime,\mu} a^{n\mu}_{\nu\nu^\prime}
\frac{(k^2_1)^{n-(\nu + \nu^\prime)+\mu}}{c_1^{n-\nu} c_2^{n-\nu^\prime}}
\frac{(k^2_2)^\mu}{c_3^\nu c_4^{\nu^\prime}}}\\
& & \frac{1}{2^{\nu+\nu^\prime -2\mu}} \left\{( k^2_1 + k^2_2	-m^2_6
)^{\nu + \nu^\prime -2\mu} - \sum_{\alpha = 1,odd}^{\nu + \nu^\prime -2\mu}
(k_1^2 + k^2_2 - m^2_6)^{\nu + \nu^\prime - 2\mu - \alpha}
(2 k_1  k_2)^{\alpha-1} \cdot c_6\right\}.
\end{eqnarray*}

Let us consider the term $(\lambda_\alpha = \nu + \nu^\prime - 2\mu - \alpha;
\alpha = 0, \cdots, \nu + \nu^\prime - 2\mu; \lambda_0 \equiv \lambda)$
\begin{eqnarray*}
(k^2_1 + k^2_2 - m^2_6)^{\lambda_\alpha}
 = \lambda_\alpha ! \sum^{\lambda_\alpha}_{\beta = 0} (-1)^\beta
\frac{(m_6)^\beta}{\beta !} \sum^{\lambda_\alpha - \beta}_{\gamma =0}
\frac{ (k^2_1)^{\lambda - \delta}}{(\lambda - \delta)!}
\frac{ (k^2_2)^\gamma}{\gamma !},
\end{eqnarray*}
where $\delta = \alpha + \beta + \gamma$ was introduced.
Accordingly we have to cancel scalar products in the following combination
\begin{equation}
\frac{(k^2_1)^{n-\mu - \delta}}{c_1^{n-\nu+1} c_2^{n-\nu^\prime +1}}
\cdot \frac{(k^2_2)^{\mu +\gamma}}{c_3^{\nu+1} c_4^{\nu^\prime +1} c_5
} ~,
\label{combi}
\end{equation}
In both factors of
(\ref{combi}) the sum of powers of $k^2_1$ and $k^2_2$, respectively, is
larger in the denominator than in the numerator so that complete cancellation
is
possible. Expanding, e.g.,
\[
(k^2_1)^{n-\mu-\delta} = \sum^{n-\mu-\delta}_{\kappa =0} {n-\mu-\delta
\choose \kappa} c_1^{n-\mu-\delta-\kappa} (m^2_1)^\kappa,
\]
we have to consider the ratio
\[
R^\kappa_1 = \frac{c_1^{\nu - \mu -\delta -\kappa -1}}{c_2^{n-\nu^\prime
+1}}\ .
\]

Depending on the sign of $\nu_1 = \nu-\mu-\delta-\kappa -1$, the partial
fraction decomposition is performed in the following manner (see also
Ref. \cite{FORM}, ch.~10):

\begin{description}
\item{1)} $\nu_1 \ge 0$: in this case we simply have
\[
R^\kappa_1 = \sum^{\nu_1}_{i=0} {\nu_1 \choose i} \frac{(m^2_2 - m^2_1)^i}
{c_2^{n-\nu^\prime - \nu_1 + 1 + i}} ~,
\]
i.e. there remains no $c_1$ in the decomposition.

\item{2)} $\nu_1 < 0$:
\begin{eqnarray*}
R^\kappa_1 &=& \sum^{|\nu_1 |-1}_{i=0} (-1)^i {n - \nu^\prime +i
\choose n-\nu^\prime} \frac{1}{(m^2_1 - m^2_2)^{n-\nu^\prime + 1 + i}
c_1^{|\nu_1 |	-i}}\\
&+& \sum^{n-\nu^\prime}_{i=0} (-1)^{|\nu_1|} {|\nu_1 |-1+i \choose
|\nu_1 |-1} \frac{1}{(m^2_1 - m^2_2)^{|\nu_1 | + i}	c_2^{n - \nu^\prime
+1-i}}\ .
\end{eqnarray*}
\end{description}

Similarly we proceed for the $k^2_2$-dependent part of (\ref{combi}). Expanding
\[
(k^2_2)^{\mu+\gamma} = \sum^{\mu +\gamma}_{\kappa=0} {\mu + \gamma \choose
\kappa} c_4^{\mu + \gamma -\kappa} (m^2_4)^\kappa\ ,
\]
we deal with
\[
R^\kappa_2 = \frac{c_4^{\mu + \gamma - \nu^\prime - \kappa -1}}
{c_3^{\nu+1}} ~,
\]
introducing $\nu_2 = \mu + \gamma - \nu^\prime - \kappa -1$ and
performing the partial fraction decomposition as above.
Since, however, $c_5 = k^2_2 - m^2_5$ is also $k_2$-dependent, for each
power of $1/c_3$ and $1/c_4$ a further decomposition has to be performed,
like e.g.
\[
\frac{1}{c_3^{p+1} c_5} = - \sum^p_{i=0} \frac{1}{(m^2_5 - m^2_3)^{p+1-i}}
\frac{1}{c_3^{i+1}} + \frac{1}{(m^2_5 - m^2_3)^{p+1}} \frac{1}{c_5}\ .
\]
For	$\alpha =0$ we obtain in this manner ``genuine" two-loop integrals
of the type (\ref{VBs}) which will be dealt with in Sect.~6, while for
$\alpha \ge 1$ the factorized one-loop integrals are of the type (see also
Ref. \cite{asy})
\[
\int \frac{d^d k_1 d^d k_2}{(k^2_1 - m^2_1)^{\nu_1}(k^2_2 - m^2_2)^{\nu_2}}
(2k_1k_2)^N = \frac{N!}{(\frac{N}{2})! {(\frac{d}{2})}_{\frac{N}{2}}}
I_{\nu_1}^{(N)} (m_1) I^{(N)}_{\nu_2} (m_2)
\]
with
\[
I^{(N)}_\nu (m) = \int \frac{d^d k}{(k^2 - m^2)^\nu} (k^2)^{\frac{N}{2}}
= i^{1-d} \pi^{\frac{d}{2}} (-m^2)^{\frac{d}{2} + \frac{N}{2} - \nu}
\frac{\Gamma (\nu - \frac{N}{2} - \frac{d}{2} )
{(\frac{d}{2})}_{\frac{N}{2}}}
{\Gamma (\nu)}\ .
\]

Somewhat differently works the cancellation of the numerator scalar products
of the integration momenta for the {\it non-planar} case since here
$k^2_3$ occurs also in $c_4 = k^2_3 - m^2_4$ which is raised to high inverse
powers. Therefore it is advisable to expand the full product of scalar
products of different momenta in (\ref{fullA}) in terms of squares of
integration momenta, i.e. we write with $\lambda_1 = \nu - \mu - \sigma
+ \tau, \lambda_2 = \nu^\prime - \mu + \sigma - \tau, \lambda_3 = \mu - \sigma
- \tau$ and $\lambda = \lambda_1 + \lambda_2 = \nu + \nu^\prime - 2\mu ,
 \Lambda = \lambda_1 + \lambda_2 + \lambda_3 =  \nu + \nu^\prime
- \mu - \sigma - \tau$
\begin{eqnarray*}
(k_1 k_2)^{\lambda_1} (k_1 k_3)^{\lambda_2} (k_2 k_3)^{\lambda_3}
&=& \frac{1}{2^\Lambda} (k^2_1 + k^2_2 - k^2_3)^{\lambda_1} (k^2_1 - k^2_2
+ k_3^2)^{\lambda_2} (k^2_1 - k^2_2 - k^2_3)^{\lambda_3}\\
&=& \frac{1}{2^\Lambda} \sum^\Lambda_{\alpha = 0} (k^2_1)^{\Lambda - \alpha}
P_\alpha (k^2_2,k^2_3)\ .
\end{eqnarray*}

A little algebra also allows to expand $P_\alpha$:
\begin{eqnarray}
P_\alpha (k^2_2, k^2_3) &=& (-1)^\alpha \sum^\alpha_{\beta=0}
(-1)^\beta f_{\alpha,\beta}
(k^2_2)^\beta (k^2_3)^{\alpha - \beta} 	\qquad \mbox{with}\\
f_{\alpha,\beta} &=& \sum^{min(\alpha,\lambda_3)}_{i=max(0,\alpha-\lambda)}
{\lambda_3 \choose i} g^i_{\alpha,\beta}
g^{\lambda_1}_{\lambda ,\alpha -i}\\
\mbox{and} & & \nonumber \\
g^k_{\alpha,\beta} &=&\sum^{min(k,\beta)}_{\ell=max(0,k-(\alpha-\beta))}
 (-1)^\ell	{\alpha-k \choose \beta-\ell} {k\choose \ell}\ .
\end{eqnarray}
$k^2_1$, e.g., can now be completely cancelled only if
\[
n-\mu \ge \Lambda = n - \mu + \theta
\]
with $\theta = \nu + \nu^\prime - n - \sigma - \tau \le 0$. Since
obviously we do not always have $\theta \le 0$ (in contrary to the planar
case with $n - (\nu + \nu^\prime) + \mu \ge 0$, see (\ref{hernja})) not all
$k^2_1$ can be cancelled, and similarly the same holds for $k^2_2$ and
$k^2_3$. This means that we will obtain integrals of the type (\ref{VBs})
with negative indices for which explicit expressions will be given at the
end of Sect.~6. In fact these integrals are also factorized one-loop integrals
which only appear in a somewhat different manner than in the planar case.

The following expansions are needed now:
\begin{eqnarray*}
(k^2_1)^{n-(\nu + \nu^\prime) + \mu + \Lambda - \alpha} &=&
\sum^{n-\sigma-\tau-\alpha}_{\kappa =0} {n-\sigma-\tau-\alpha \choose
\kappa} c_1^{n-\sigma-\tau-\alpha - \kappa} (m^2_1)^\kappa\\
(k^2_2)^{\sigma + \beta} &=& \sum^{\sigma + \beta}_{\kappa =0}
{\sigma + \beta \choose	\kappa} c_3^{\sigma + \beta - \kappa}
(m^2_3)^\kappa \quad \mbox{and}\\
(k^2_3)^{\tau + \alpha - \beta} &=& \sum^{\tau + \alpha - \beta}_{\kappa
=0} {\tau + \alpha - \beta  \choose \kappa} c_4^{\tau + \alpha - \beta
- \kappa} (m^2_4)^\kappa
\end{eqnarray*}
and partial fraction decompositions must accordingly be performed for the
ratios:
\[
\frac{c_1^{\nu-\sigma-\tau-\alpha-\kappa-1}}{
c_2^{n-\nu^\prime +1}} \ ,
\frac{c_3^{\sigma + \beta -\nu - \kappa-1}}{c_5} \quad
\mbox{and} \quad \frac{c_4^{\tau + \alpha - \beta - \nu^\prime -
\kappa-1}}{c_6}\ .
\]

This works out in analogy to the planar case and will not be discussed in
further details.

\vglue 0.6cm
{\elevenbf\noindent 6. Recurrence relations}
\vglue 0.2cm

   As was discussed in Sect.2, the final step of our algorithm
consists in the evaluation of the bubble integrals (\ref{VBs})
after the Taylor coefficients are finally expressed in terms of these.
Their resolution is supposed to be performed in terms of recurrence
relations.
These were first considered
in \cite{Hoog}. One can get such relations from the identity:

\vskip 0.5cm
\begin{equation}
\int \frac{d^d~k_1~d^d~k_2}{[k_2^2-m_2^2]^{\beta} }
{}~~ \frac{\partial}{\partial k_{1 \mu}}
\left( \frac{A~ k_{1 \mu}-~B~k_{2\mu}}
{[k_1^2 -m_1^2]^{\alpha} [(k_1-k_2)^2-m_3^2]^
 {\gamma }} \right) \equiv 0
\end{equation}
with arbitrary constants $A$ and $B$.
Taking, for example, $A=(m_1^2+m_2^2-m_3^2)/2/m_1^2 B$
we obtain a recurrence relation for $V_B(\alpha,\beta,\gamma+1)$
and $V_B$'s with their sum of indices equal to
$\alpha+\beta+\gamma$.
For details on recurrence relations for Feynman diagrams
see also Refs. \cite{Chet}, \cite{Brod}.
Explicitly we obtained the following relations for two-loop
bubble diagrams with three masses (see \cite{lTaH}):

\ba
V_B&=&\frac{{\bf 1^-}}{2(j_1-1)m_1^2}\left\{j_3
    \left[{\bf 1^-}-{\bf 2^-}-(m_1^2-m_2^2+m_3^2)\right]{\bf 3^+}
				      \right.  \label{rec1} \\
\crn
&+&\left.(2(j_1-1)+j_3-d)\right\}V_B \crn
\crn
V_B&=&\frac{\Delta(m_1,m_2,m_3)}{j_2-1}                    \left\{2j_3m_3^2
    \;{\bf 2^-}\left({\bf 2^-}-{\bf 1^-}\right){\bf 3^+}
\right.\label{rec2} \\
\crn
&+&\left.(j_2-1)(m_1^2-m_2^2-m_3^2)\left({\bf 3^-}-{\bf 1^-}\right)
   \right.\crn
\crn
&+&\left.\left[ 2(j_2-1)m_3^2+2j_3(m_1^2-m_2^2)+(j_2-1-d)(m_1^2-m_2^2+m_3^2)
	   \right]{\bf 2^-} \right\}V_B \crn
\crn
V_B&=&\frac{\Delta(m_1,m_2,m_3)}{j_3-1}                    \left\{2j_1m_1^2
    \;{\bf 1^+}\left({\bf 3^-}-{\bf 2^-}\right){\bf 3^-}
\right.\label{rec3} \\
\crn
&+&\left.(j_3-1)(m_1^2-m_2^2+m_3^2)\left({\bf 2^-}-{\bf 1^-}\right)
   \right.\crn
\crn
&+&\left.\left[ 2(j_3-1)m_1^2+2j_1(m_2^2-m_3^2)+(j_3-1-d)(m_1^2+m_2^2-m_3^2)
	   \right]{\bf 3^-} \right\}V_B, \nonumber
\ea
where ${\bf 1^{\pm}}V_B(j_1, \dots, m_1, \dots )
		   \equiv V_B(j_1{\pm} 1, \dots, m_1, \dots )$ etc.~
and
\begin{equation}
\Delta(m_1,m_2,m_3) = \frac{1}{2m_1^2m_2^2+2m_1^2m_3^2+2m_2^2m_3^2
				 -m_1^4-m_2^4-m_3^4}~~~~.
\end{equation}
The idea of their application is to reduce all integrals to the master
integral
\begin{equation}
V_B(1,1,1,m_1,m_2,m_3)
\label{master}
\end{equation}
and some simple tadpole-integrals, which are obtained when one of the
indices is zero.
By inspection one observes that applying all three recurrence relations
one
after the other, the first index does not increase and the sum of all
indices decreases by at least 2 (by 1 in each of the last two steps).
In
this manner $j_2$ or $j_3$ must get zero and the procedure can be stopped.\\
 A particular point we have to observe is that the
integrals (\ref{VBs}) may diverge, i.e. writing the space-time
dimension as $d=4-2\varepsilon$, we will have poles in $\varepsilon$~.
Since we intend to develop our algorithm in such a manner that it
can be implemented in terms of a FORTRAN program we have to
take special care of these poles, i.e. we have to split the integrals
(\ref{VBs}) into their finite and their divergent part. The
master integral (\ref{master}) and $V_B(1,1,2,m_1,m_2,m_3)$ are the
only ones which have a pole of
second order (for details see also \cite{ft}), and only the integrals
$V_B(1,1,n,m_1,m_2,m_3)$ with n $>$ 2 have poles of first order, all
others being finite. Thus we write
\begin{equation}
V_B(1,1,n,m_1,m_2,m_3)=F(1,1,n,m_1,m_2,m_3)+\frac{1}{\varepsilon}~
I~(1,1,n,m_1,m_2,m_3), ~n > 2 .
\end{equation}

In what follows we will show a path of how to resolve equations
(\ref{rec1}) - (\ref{rec3}).
For convenience we drop the masses in the argument list.
Using recurrence equation (\ref{rec3}) we get
\begin{eqnarray*}
&&V_B(1,1,n) ~~= \\
&& \frac{\Delta}{n-1}\left\{ 2m^2_1\left[V_B(2,1,n-2)-V_B(2,0,n-1)\right]
 + \right. \\
	       &&   (n-1)(m^2_1 - m^2_2 + m^2_3)
               \left[V_B(1,0,n)-V_B(0,1,n)\right] +  \\
	       &&  \left.  \left[2(n-1)m^2_1 + 2(m^2_2 - m^2_3)
	       + (n-1-d)(m^2_1 + m_2^2
	       - m^2_3)\right]V_B(1,1,n-1)\right\}.
\end{eqnarray*}

Here $V_B(2,0,n-1), V_B(1,0,n)$ and $V_B(0,1,n)$ are ``trivial''
and will be substituted explicitly, e.g.
\begin{equation}
V_B(0,m,n) =   (-1)^{m+n}
                   \frac{\Gamma \left (m-\frac{d}{2} \right)}{\Gamma
					 \left(m\right)}
		   \frac{i \pi^{n/2}}{ {m_2^2}^{\left( m-\frac{d}{2}
						\right )}}
		   \frac{\Gamma \left (n-\frac{d}{2} \right)}{\Gamma
						\left(n\right)}
		   \frac{i \pi^{n/2}}{ {m_3^2}^{\left( n-\frac{d}{2}
						\right )} }
\end{equation}
with
\begin{equation}
\Gamma (1-\frac{d}{2})= -\frac{1}{\varepsilon}-1-\varepsilon.
\end{equation}
This indicates the occurrence of a pole in $\varepsilon$ which gives a
contribution to I(1,1,n). $V_B(2,1,n-2)$ is
found by an application of recurrence equation (\ref{rec1}):
\begin{eqnarray*}
V_B(2,1,n-2) &=& \frac{1}{2m^2_1} \left\{ (n-2)\left[V_B(0,1,n-1)-V_B(1,0,n-1)
  - \right. \right. \\
& & \left. \left.  (m^2_1 - m^2_2 + m_3^2) V_B(1,1,n-1)\right] + (n-d)
V_B(1,1,n-2)
\right\} .
\end{eqnarray*}
Therefore we can inductively calculate $V_B(1,1,n)$ for all $n \ge 3$ using
the master integral $V_B(1,1,1)$. Care has to be taken in evaluating
terms of the form $d~V_B(1,1,n-1)$ on the r.h.s. which yield a finite
and an infinite part with $d = 4 - 2 \varepsilon $ .
Using recurrence equation (\ref{rec2}), we get moreover
\begin{eqnarray*}
&&V_B(1,m,n)= \\
&&\frac{\Delta}{m-1}\left\{ 2nm^2_3 \left[V_B(1,m-2,n+1)-V_B(0,m-1,n+1)\right]
	       +\right. \\
	       &&  (m-1)(m^2_1 - m^2_2 - m^2_3) \left[V_B(1,m,n-1) -
	       V_B(0,m,n)\right] + \\
	       && \left. \left[2(m-1)m^2_3 + 2n(m^2_1 - m^2_2) + (m-1-d)(m^2_1 -
	       m_2^2	+ m^2_3)\right] V_B(1,m-1,n)\right\} \
\end{eqnarray*}
so that	$V_B(1,m,n)$ is computable, and finally an application of recurrence
equation (\ref{rec1}) leads to
\begin{eqnarray*}
&&V_B(l,m,n)=\\
&& \frac{1}{2m^2_1 (l-1)} \left\{ n\left[V_B(l-2,m,n+1) - V_B(l-1,m-1,n+1)-
 \right. \right. \\
&& \left. \left. (m^2_1 - m^2_2 + m_3^2) V_B(l - 1, m, n+1)\right] + (2(l-1) +
n-d)
V_B(l-1,m,n)\right\}\ .
\end{eqnarray*}

Implementing the recurrence relations numerically, it is advisable,
in order to take
properly into account the poles in $\varepsilon$, to produce explicit
relations for the lower indices by means of FORM. For the
$V_B(1,1,n)$, e.g., the $\frac{1}{{\varepsilon}^2}$ pole from the
master integral causes ``extra'' divergences and for the $V_B(l,m,n)$
divergences come in for lower indices whenever a $V_B(1,1,n)$ is
encountered in a recurrence relation while for higher indices these
integrals are finite as can also directly be seen from (\ref{VBs}).

   Apart from that, the recurrence relations are implemented in
exactly the order as described above: first $V_B(1,1,n)$ for all $n$,
secondly $V_B(1,m,n)$ and finally $V_B(l,m,n)$. In each of these cases
the recurrence has been implemented for all $l + m + n \leq J $
($l,m,n \geq 1$; for negative indices see below), where
the large integer $J$ is to be chosen according to the number of
Taylor coefficients needed.

   Of course one may have doubts about the numerical stability of such
a recursive approach, in particular if one knows that indeed the
Taylor coefficients must be calculated with high precision as
mentioned in Sect.~2. Ordinary FORTRAN double precision will clearly
not do the job. We used the multiple precision package written by
D.H.Bayley \cite{Bail}, which also provides an automatic translator
for any FORTRAN program. The requested precision is here defined at the
beginning of the program. With this package we used for $J=62$ and 100
decimals precision 42 seconds on a HP735 (49 seconds for 150 decimals
precision). The numerical results were tested against numerics
performed with REDUCE for explicit expressions of the integrals in
the equal mass case. In principle, however, the best precision test
will always be to increase the number of decimals used in the
calculation. In this manner we can be sure to have an effective
algorithm for the calculation of two-loop integrals.

   To conclude this sections we give an explicit formula for
$V_B(\alpha , \beta , \gamma)$ for the case that one of the indices is
negative. In this case the $V_B$'s can be reduced to factorized
one-loop integrals, which makes the application of the above recurrence
relations superfluous.

\begin{eqnarray}
&&V_B{(\alpha, \beta, \gamma)}=(-1)^{(\alpha+ \beta+ \gamma)}
\int \frac{d^d k_1~ d^d k_2~ [(k_1-k_2)^2-m_3^2]^{|\gamma|}}
 {(k_1^2-m_1^2)^{\alpha} (k_2^2-m_2^2)^{\beta}}=\\*
&& \nonumber \\*
&&(i \pi^{\frac{d}{2}})^2 |\gamma|!
 \sum_{l=0}^{[|\gamma|/2]} \frac{{(\frac{d}{2})}_l}{l!}
 \sum_{r=max(0,2l+\alpha-|\gamma|-1)}^{\alpha-1}\sum_{q=0}^{\beta-1}
\frac{\Gamma(1+r-l-\frac{d}{2})
\Gamma(1+q-l-\frac{d}{2})}
 {r! q! (\alpha-1-r)! (\beta-1-q)!}  \nonumber\\*
&& \nonumber \\*
&&~~~~~~~~~~~~~~~~
 \frac{s^{|\gamma| -2l-\alpha-\beta+r+q+2} (m_1^2)^{(\frac{d}{2}+l-r-1)}
 (m_2^2)^{(\frac{d}{2}+l-q-1)} }
 {(|\gamma| -2l-\alpha-\beta +r+q+2)!}~,
 \nonumber
\end{eqnarray}
where $\gamma < 0$ and $s=-m_1^2-m_2^2+m_3^2.$

\vskip 1.0cm

\newpage

\vglue 0.6cm
{\elevenbf\noindent 7. Conclusions}
\vglue 0.4cm

 An effective method to calculate Feynman diagrams has been
developed in \cite{ft}. In the present work we have been able to work
out details of an algorithm, which will finally allow to elaborate the
above method into a ``package'' for the evaluation of two-loop
three-point functions and possibly beyond. The essential new points
worked out here are described in Sects. 4 and 6: deriving an explicit
formula for the numerators in the integral representation of the
Feynman diagrams's Taylor coefficients and demonstrating the
possibility to evaluate the recursion relations for the bubble
integrals numerically by means of the multiple precision package of
\cite{Bail}. In the development of this algorithm FORM \cite{FORM}
has been the main tool in the formulae manipulation.

\bigskip
\noindent
{\elevenbf \noindent Acknowledgments}
\vglue 0.2cm

 We are very much indebted to J. Vermaseren to indicate to us on this
conference the new possibility of FORM, namely the evaluation of the
totally symmetric tensor as special function, which enabled us to
develop our algorithm in the presented form. We are also grateful to
W. Koepf for discussions on the analytic resolution of the recurrence
relations by means of non-commutative Gr\"obner basis techniques.
Thanks as well to L. Avdeev for helpfull discussions.


\bigskip
\vglue 0.5cm



\end{document}